\documentclass[doublecol]{epl2} 
\usepackage[pdftex,colorlinks=true,pdfstartview=Fit]{hyperref}		

\usepackage[normalem]{ulem}		

\usepackage{color}			

\usepackage{bm}		
\usepackage{amssymb,amsmath}
\usepackage{graphicx}

\title{Arbitrarily slow, non-quasistatic, isothermal transformations}

\author{Mom\v cilo Gavrilov\inst{1} \and John Bechhoefer\inst{1}}
\shortauthor{M. Gavrilov and J. Bechhoefer}

\institute{                    
  \inst{1} Dept. of Physics, Simon Fraser University, Burnaby, B.C., V5A 1S6, Canada
}


\abstract{
For an overdamped colloidal particle diffusing in a fluid in a controllable, virtual potential, we show that arbitrarily slow transformations,  produced by smooth deformations of a double-well potential, need not be reversible.  The arbitrarily slow transformations do need to be fast compared to the barrier crossing time, but that time can be extremely long. We consider two types of cyclic, isothermal transformations of a double-well potential.  Both start and end in the same equilibrium state, and both use the same basic operations---but in different order.  By measuring the work for finite cycle times and extrapolating to infinite times, we found that one transformation required no work, while the other required a finite amount of work, no matter how slowly it was carried out.  The difference traces back to the observation that when time is reversed, the two protocols have different outcomes, when carried out arbitrarily slowly.  A recently derived formula relating work production to the relative entropy of forward and backward path probabilities predicts the observed work average.
}

\begin{document}

\maketitle

\section{Introduction}
Both theoretical and technical advances in the past two decades have made it possible to investigate the thermodynamics of a colloidal particle in potential wells whose shape can be accurately controlled.  These advances include the precise control of the shape of a potential\cite{blickle06, cohen05d} and the ability to estimate work from potential shape and particle trajectory \cite{sekimoto97,sekimoto10}.  A colloidal particle, therefore, becomes a model system for testing experimentally many concepts in thermodynamics that heretofore have been only thought experiments.  Recent achievements include the design of a Szilard engine \cite{leff03, toyabe10} that converts information to work, and tests of Landauer's erasure principle\cite{berut12,Jun14} which states that erasing one bit of memory requires an average work of at least $kT\ln$2.  Another  experimental study interprets memory erasure as the restoration of a broken symmetry and investigates, more generally, how sudden changes in ergodicity affect the thermodynamics of a colloidal bead \cite{roldan14}. 

Here, we investigate cyclic transformations of a double-well potential, where we slowly and smoothly vary the barrier height and asymmetry of left and right scales of the potential.  Our system, a colloidal bead in a fluid, starts and ends in the same global equilibrium state in a symmetric double-well potential, with the barrier crossing time much longer than the duration of any transformation.  All transformations are isothermal, in contact with a single heat bath.

We estimate work in an arbitrarily slow limit, by extrapolating work measurements to infinite cycle times. The arbitrarily slow limit is still faster than the barrier crossing time, but that time can be longer than the lifetime of the Universe.  Naively, one might think that an isothermal, continuous, arbitrarily slow cyclic manipulation of a potential will require no work over one cycle.  Nonetheless, we will see that, for some protocols, the average work in such a limit can be positive.

\begin{figure}
	 \begin{center}
	 \includegraphics[width=8cm]{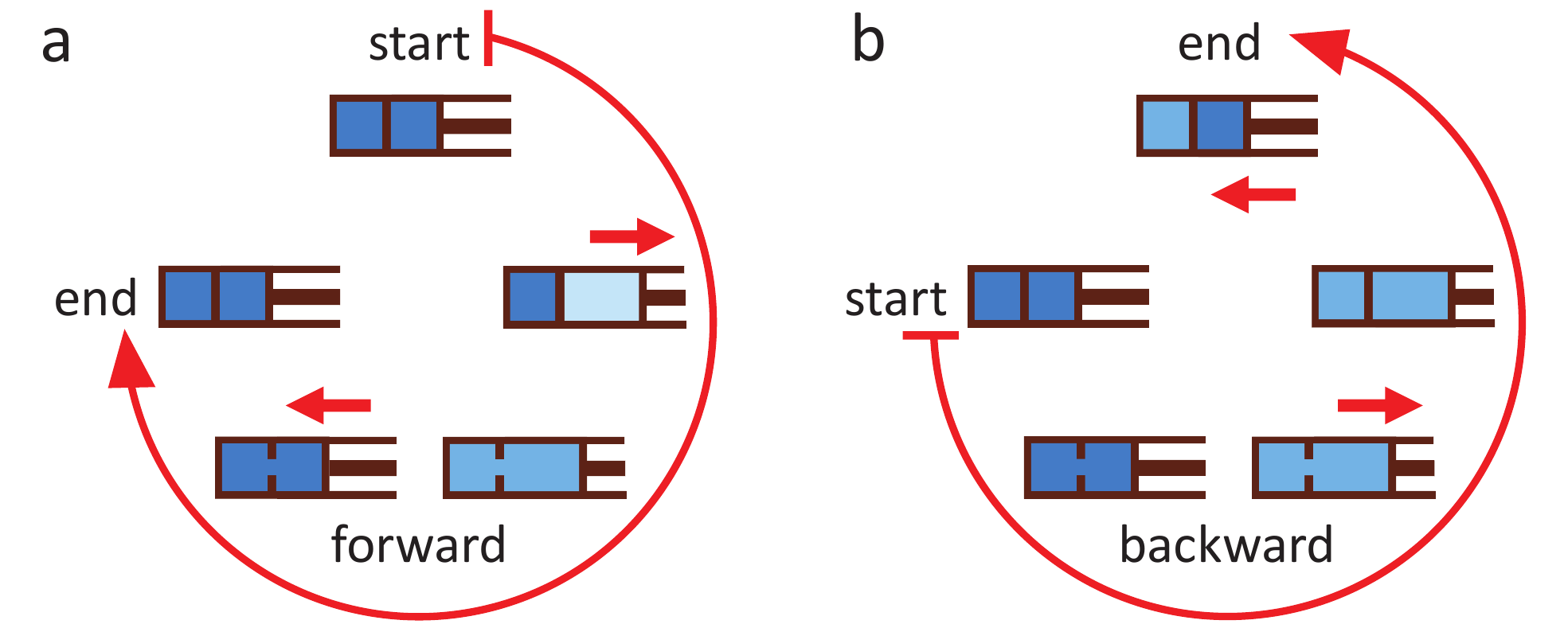}
	 \end{center}
	 \caption { \label{fig:gasSimple} 
(Color online)  
Ideal gas in a closed container with two chambers.  Valve on divider can open to equalize pressures.  Piston can adjust volume of right chamber.  (a) forward and (b) backward cycles.  Deeper shading denotes higher pressure.
}
\end{figure}

To introduce the phenomenon in a familiar setting, consider an ideal gas in a closed container with two chambers (fig.~\ref{fig:gasSimple}).  The container has diathermal walls and is connected to a heat bath (not shown).  Volumes are large enough that fluctuations may be neglected.  In fig.~\ref{fig:gasSimple}(a), the cycle starts in global thermodynamic equilibrium, with the two chambers thermally linked but separated by a closed valve.  In the next stage, moving the piston to the right expands the right-hand chamber, decreasing its pressure.  In the third stage, the valve is opened, and the pressures equalize, an irreversible step similar to free expansion.   In the fourth stage, the piston compresses the right-hand chamber back to its original size.  In the final stage, the valve is closed, returning the system to its starting equilibrium state.  All processes are slow enough to be isothermal.

Figure~\ref{fig:gasSimple}(b) shows the cycle in reverse.  Notice that in the final stage, the two pressures are \textit{different}, as denoted by the deeper shading of the right-hand chamber.  The difference between forward and reverse cycles at the conclusion reveals the nonequilibrium nature of the cycle.  When the valve is closed, there are, in effect, two separate subsystems.  Even when each is in equilibrium with respect to its local constraints, the global system can be out of equilibrium.

In this paper, we describe mesoscopic experiments that show a similar phenomenon in a single-particle setting that is both subtle and surprising:  First, mesoscopic experiments can realize equivalents to ``textbook" thought experiments such as those described above, with a precision $ \ll kT$.  Second, unlike the macroscopic case, observing one cycle is not enough to distinguish equilibrium from nonequilibrium cases.  A statistical analysis is needed.  Third, the nonequilibrium, irreversible effects will be observed in a smoothly varying potential that is manipulated arbitrarily slowly.

\section{Feedback trap}
A feedback trap is a technique for trapping and manipulating small particles and molecules in solution.  Feedback traps have been used to measure physical and chemical properties of particles \cite{cohen06a,cohen05d,wang14,cohen07a,Goldsmith2010} and to explore fundamental questions in the nonequilibrium statistical mechanics of small systems \cite{cohen05b,cohen05d,jun12,gavrilov13}.  For the latter application, one can impose a trapping potential $U(x, y, t)$ and explore a particle's dynamics in it.  The feedback trap creates a \textit{virtual potential} by rapidly cycling through three steps:  
\begin{itemize}
\item observe the position $(\bar{x}_n,\bar{y}_n)$ of a trapped object;
\item calculate a force from the imposed potential at the observed position $\vec{F}_n=-\triangledown U|_{\bar{x},\bar{y},n\Delta t}$;
\item  apply a voltage proportional to the calculated 2D force to four electrodes that create a horizontal field in the $xy$-plane.
\end{itemize}
The field is kept constant until updated with information acquired from the next image.  

A feedback trap can impose any two-dimensional potential, as long as the feedback loop time $\Delta t$ is short enough compared to the relaxation time for movements of the particle in the potential, $t_r$.  Since there is no physical potential trapping the particle---only the approximation to one imposed by the rapid feedback loop---the potential is \textit{virtual}.  Here, we measure the dynamics of a particle trapped in a time-dependent, double-well, virtual potential.

\section{Experimental setup}
\begin{figure}[t]
	 \begin{center}
	 \includegraphics[width=8cm]{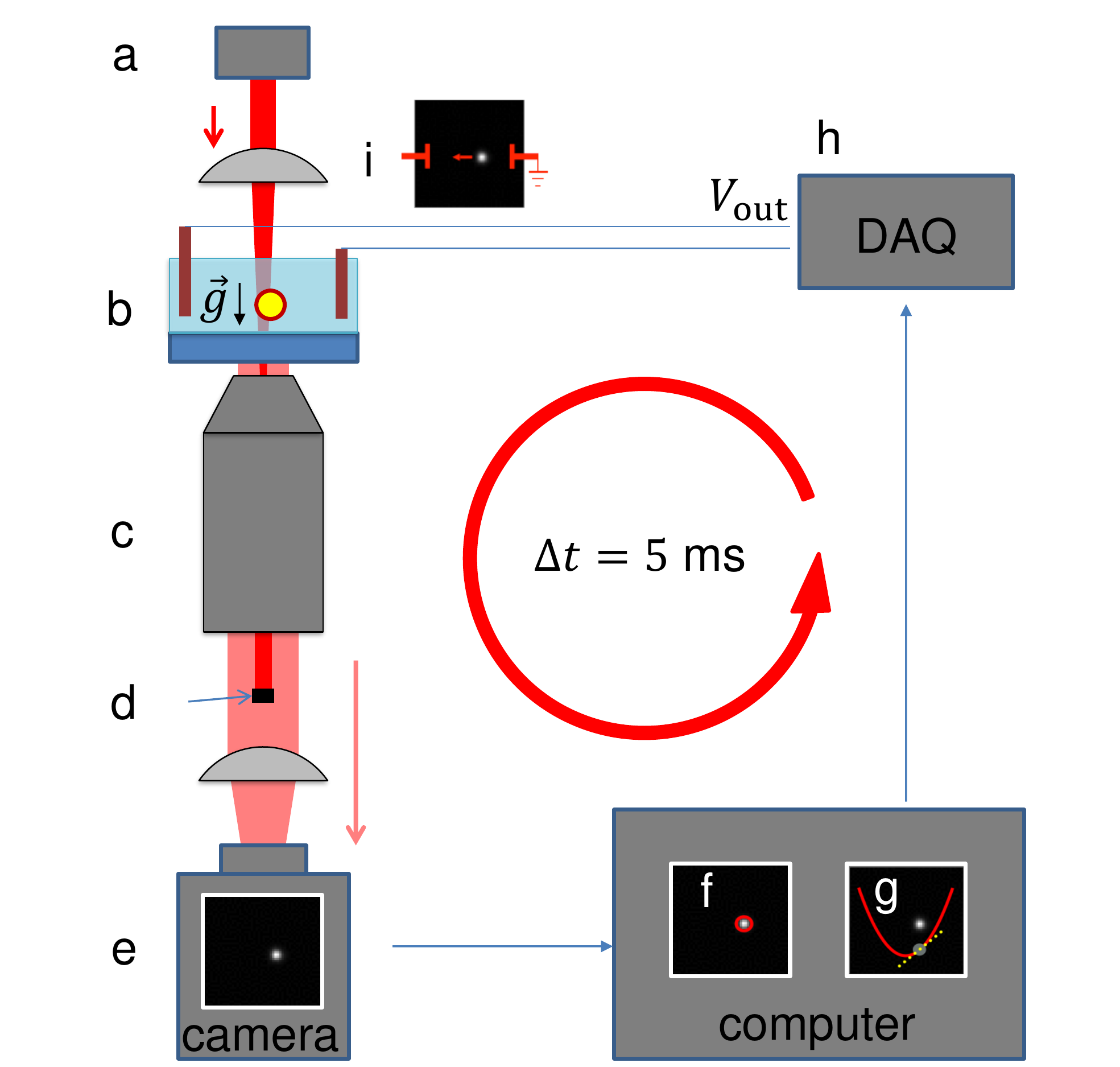}
	 \end{center}
	 \caption[example] { \label{fig:FBtrap} 
(Color online)  Schematic of feedback trap operation.  (a)  LED light source.  (b) Bead in sample cell sinks under gravity and diffuses predominantly in the XY plane.  (c) High-NA microscope objective.  (d) Beam blocker stops the LED beam, allowing only scattered light from bead to reach the camera.  (e)  Camera takes image.  (f) Image-processing program estimates bead position and (g) calculates the force based by the imposed potential.  (h) DAQ applies voltage proportional to calculated force to electrodes.  (i) Electric and thermal forces move bead to a new position.
The feedback loop repeats indefinitely, with cycle time $\Delta t = 5$ ms.}
\end{figure}

We trap an overdamped silica bead of diameter 1.5~$\mu$m in a virtual potential created by a feedback loop with update time $\Delta t = 5$ ms (fig.~\ref{fig:FBtrap}).  The apparatus is based on a home-built dark-field microscope equipped with a 60x Olympus NA=0.95 air objective \cite{weigel14,gavrilov15}.  An LED source (660~nm, Thorlabs) illuminates the sample, and a camera (Andor iXon DV-885) takes a $50 \times 20$ pixel image every $\Delta t$~=~5~ms, with an exposure $t_c$~=~0.5~ms.  The potential is a double well along the $x$ axis and harmonic along the $y$ axis.  A LabVIEW program processes acquired images by thresholding the background and applying a modified centroid algorithm to a small region of interest centered on an initial guess (the intensity maximum) \cite{berglund08,gavrilov13}.  

Based on the estimated position and the virtual potential, voltages are applied at a time $t_d$~=~5~ms after the middle of the exposure.  The DAQ voltages are amplified 15x using a custom-built amplifier and held constant for 5~ms, until the next update. 
The working particles in our trap were chosen to be large enough and heavy enough that gravity confines their motion to within the depth of focus of our microscope, yet  small enough to diffuse freely in the horizontal plane.  In particular, the particles had a diameter $2a =1.5~\mu$m and were made of silica (density $\rho_s = 2.2$ gm/cc, in water).

The key technical issue in feedback traps of this type is to ensure that the force applied to the particle accurately approximates forces from the desired potential.  The calibration procedure \cite{gavrilov14} leads to running estimates of the particle's mobility matrix $\boldsymbol{\mu}$ (linear response of velocity to applied voltage) and diffusion constant $D$.  With a calibrated trap, one can estimate the work during a time-dependent protocol.

\section{Virtual potential}
Using the feedback trap, we impose a time-dependent, double-well potential, parametrized for independent control of the barrier height and width of each well.  The two-dimensional virtual potential has the form 
\begin{equation}
	U(x,y,t) = E_b \left[ -2 g(t) \left (\frac{x}{x_\eta(t)} \right)^2 + \left( \frac{x}{x_\eta(t)} \right)^4 \right]
		+\frac{1}{2} \kappa y^2 \,,
\label{eq:DWpotential}
\end{equation}
where the functions $x_\eta(t)$ and  $g(t)$ define the experimental protocols.  The maximal barrier height $E_b$ and the stiffness $\kappa$ of the $y$ axis harmonic potential are constant.  

\begin{figure}[tb]
	 \begin{center}
	 \includegraphics[width=8cm]{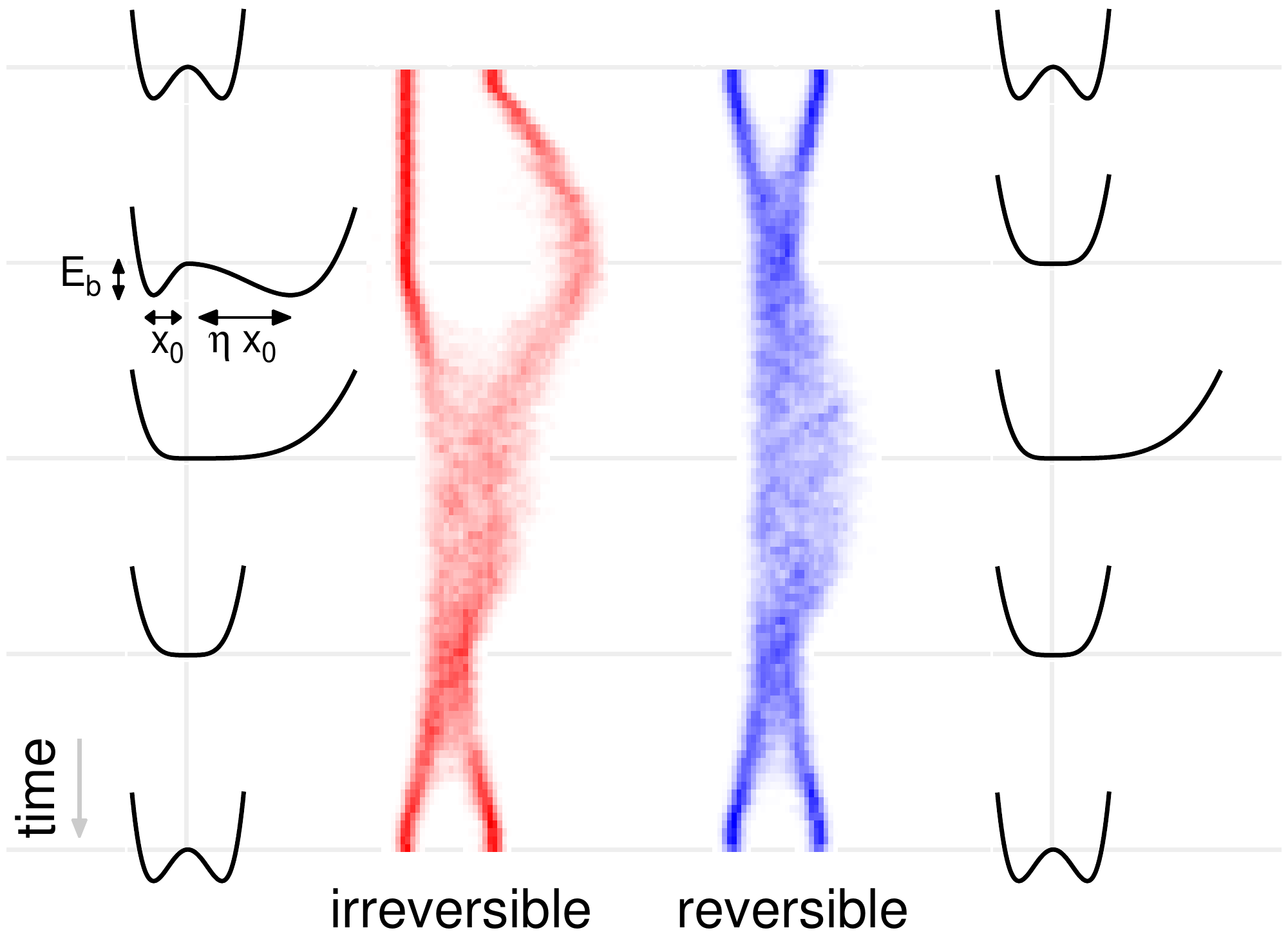}
	 \end{center}
	 \caption { \label{fig:trajectories} 
(Color online)  Transformation protocols and time evolution of probability for fixed cycle time.  The images are two-dimensional histograms, with intensity $\propto P(x,t)$ the occupation probability for a particle in a time-dependent, double-well potential.  Each histogram was generated from 60 trajectories.  The scale parameters are barrier height $E_b = 13~kT$, well location $x_0 =  0.77~\mu$m, and asymmetry $\eta = 3$.
}
\end{figure}

In the double-well potential, the function $x_\eta(t)$ controls the asymmetry by taking different values on the left and right sides of the potential:
\begin{equation}
x_\eta(t) = 
	\begin{cases}
		x_0  &  x<0 \\
		\left[ 1+ f(t)(\eta-1) \right] x_0  &  x\geq 0 \,,
	\end{cases}
\label{eq:Asymmetry}
\end{equation}
where $f(t) \in [0,1]$ acts as a stretching protocol, adjusting the distance between the potential minimum and the origin, $x=0$.  For $f = 0$, the potential is symmetric, with both wells at a distance $x_0$ from the origin.  For $f = 1$, the width of the right well is stretched by a factor of $\eta$ relative to the left.  The asymmetry factor $\eta \in [1,4]$.  The function $g(t) \in [0,1]$ in eq.~\ref{eq:DWpotential} controls the barrier height, with the maximal barrier height $E_b$ corresponding to $g(t)=1$ and no barrier corresponding to $g(t)=0$.  The protocols shown in fig.~\ref{fig:trajectories} illustrate the manipulation of barrier heights and stretching of wells.  The work to manipulate a potential in one cycle of duration $t_{\rm cyc}$ is estimated by discretizing  Sekimoto's formula \cite{sekimoto97,sekimoto10} for the stochastic work $W_{\rm cyc} = \int_0^{t_{\rm cyc}} dt \, \left. \tfrac{\partial U(x,t)}{\partial t} \right|_{x=x(t)}$.

\section{Arbitrarily slow limit}

To understand this limit, we consider the time scales in our problem.  
As discussed previously, the smallest timescale is the update time $\Delta t$~=~5~ms, set by the hardware of our experiment.  Relative to $\Delta t$, we choose the potential so that the particle relaxation time $t_r$ within the basins is much greater than the update time $\Delta t$.  This minimum relaxation time is calculated from the maximum curvature of the potential.  For the potential in eq.~\ref{eq:DWpotential}, the relaxation time is $t_r = x_0^2/(8E_b D)$, where $D$ is the particle diffusivity \cite{Jun14}.

A natural scale time for cyclic transformations is given by the time for a particle to freely diffuse a distance equivalent to the well separation $(1+\eta) x_0$.  Once the barrier has been lowered,  the particle will explore the entire extent of the potential in this amount of time, which is given approximately by $\tau_0~=~[(1+~\eta)x_0]^2/D$.  The dimensionless cycle time is then defined as $\tau \equiv t_{\rm cyc} / \tau_0$.  Cyclic quantities such as $W_{\rm cyc}$ are then rescaled to be $W_\tau$.

The longest timescale is the Kramers time to hop over the barrier, $t_{\rm hop}\approx\tau_0 \left( \frac{\sqrt{2}\pi}{16}\right) \left( \frac{\exp{(E_b/kT)}}{E_b/kT} \right)$ \cite{Jun14}.   This hop time can exceed the age of the Universe for modest barrier heights:  it is impossible, \textit{even in principle}, to wait ``long enough" to equilibrate the two wells.  Thus, in the arbitrarily slow limit, the dimensionless cycle time $\tau \gg 1$, and the effectively infinite hop time of the barrier creates an \textit{internal constraint} analogous to the dividing wall separating the gases in fig.~\ref{fig:gasSimple} \cite{callen85,gould10}.

\section{Experimental parameters}
For a 1.5-$\mu$m-diameter bead, typical mobility values in our sample cell are $\mu \approx$ 0.1$\mu$m/V/s.  The diffusivity $D\approx0.23~\mu$m$^2$/s, which is 0.67 times the Stokes-Einstein value of 0.35~$\mu$m$^2$/s for a similar sphere far from boundaries and consistent with the increased drag expected for a sphere near a boundary \cite{happel83}.  

For a virtual potential, the feedback loop update time $\Delta t$ should be fast compared to the (overdamped)  local relaxation time $t_r$ of particles fluctuating about the local potential minimum at $\pm x_0$ \cite{jun12}.  To satisfy this constraint, we select $t_r>\Delta t/0.2$ and further choose  $x_0^2 = 8(E_b/kT)D t_r>8(E_b/kT)D \Delta t/0.2$, where $E_b$ is the barrier height \cite{Jun14}.  Using the typical value for $D$, we choose $x_0 =  0.77~\mu$m and an energy barrier $E_b/kT = 13$, implying $\tau_0 \approx 10.3$ s and a barrier-hop time $t_{\rm hop} \approx 1$~day.  (A barrier height $\approx3\times$ larger, $44~kT$, would imply $t_{\rm hop} \approx 10^{18}$~s, twice the age of the Universe.)  By contrast, the longest cycle time was 10 min ($\tau = 50$).

\section{Results}
We analyze cyclic protocols that start from a symmetric double-well potential (fig.~\ref{fig:trajectories}, top).  In all cases, the system starts and ends in equilibrium, with equal probability for a particle to be in the left or right wells.  In practice, since we know and control the initial state of the particle (which well it is in), we can compute thermodynamic quantities such as work by simply averaging over the quantity conditioned on starting in the left well or starting in the right well.

We then analyze the two protocols shown in fig.~\ref{fig:trajectories}.  In the  \textit{irreversible} protocol (fig.~\ref{fig:trajectories}, left), we increase the asymmetry to a chosen value $\eta$, then lower the relative barrier height $E_b$, then remove the asymmetry $\eta$, and finally raise back the barrier $E_b$.  The \textit{reversible} protocol (fig.~\ref{fig:trajectories}, right) differs only in reversing the first two operations:  First, lower the barrier $E_b$; then increase the asymmetry $\eta$.  The last two operations are the same:  decrease the asymmetry $\eta$ and then raise back the barrier $E_b$. In \textit{both} protocols, particles have an equal probability of ending up in either the left or the right wells.  That is, the system (the Brownian particle in double-well potential) starts and ends in thermodynamic equilibrium with the surrounding fluid bath.  In contrast to recent work that achieves adiabatic transformations while varying the temperature \cite{martinez15,martinez15b}, the processes here are isothermal.

\subsection{Data Acquisition}
After choosing the protocol type,  asymmetry, and cycle time, we take data for about 30 minutes.  Then we proceed to a different parameter set.  Before each cycle, we set the initial condition by imposing a stiff harmonic potential in $x$ and $y$.  The harmonic trap is centered on $+x_0$ for half the measurements and on $-x_0$ for the other half.  We record the particle trajectory $x(t)$ and measure the work $W_\tau$ for each cycle.
\begin{figure*}[tbh]
	 \begin{center}
	 \includegraphics[width=16cm]{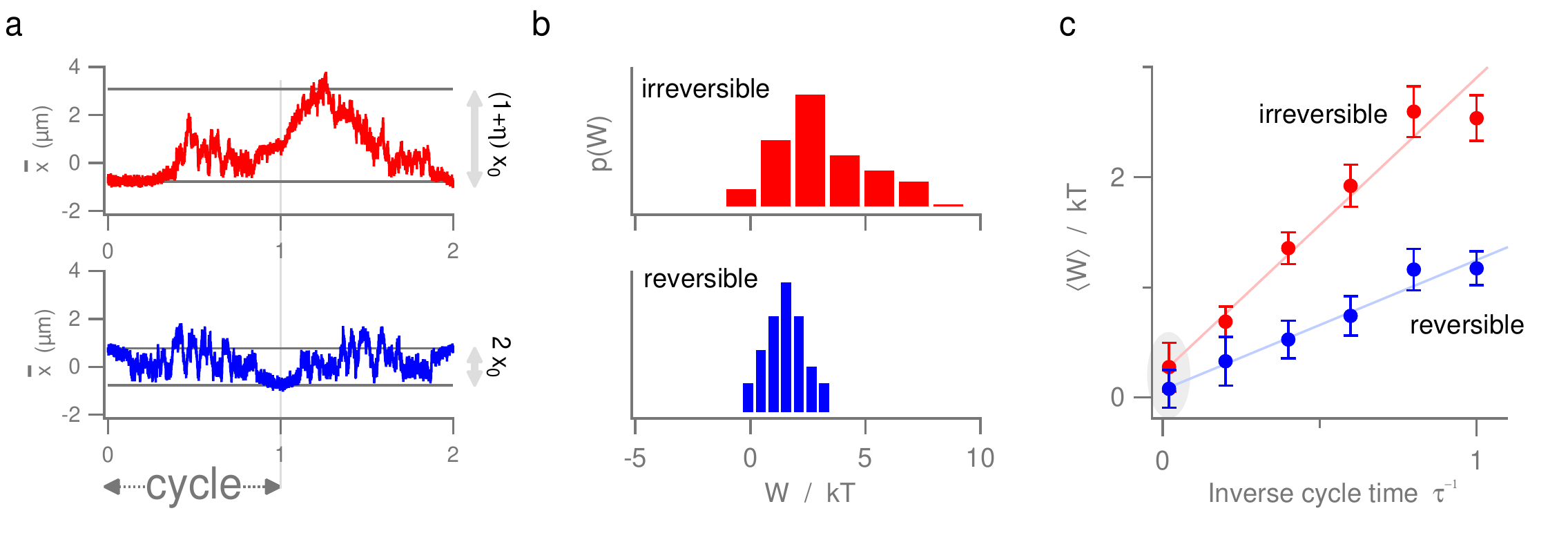}
	 \end{center}
	 \caption { \label{fig:workRIR} 
(Color online)  Work for slow transformations, in the reversible and irreversible protocols.   (a) Two trajectories of a particle for irreversible and reversible protocols ($\tau=1.25$).  (b) Work distributions $p(W)$ for fixed cycle time $\tau$, compiled from many trajectories for both protocols. (c)  Average work as a function of inverse cycle time for a fixed asymmetry factor, $\eta = 4$.  Error bars represent the standard error of the mean.  The asymptotic work $\left<W\right>_{\infty}$ corresponds to the $y$-axis intercept (shaded area).  Solid lines are fits to the experimental data.}
\end{figure*} 
\subsection{Analysis}
Figure~\ref{fig:workRIR} illustrates average work estimates for arbitrarily slow protocols.   Part (a) shows a position time series for a particle that executes irreversible (red) and reversible (blue) protocols for a cycle time $\tau =1.25$.  The histogram of position time series (fig. \ref{fig:trajectories}) is accumulated from repeated reversible and irreversible transformations at a fixed cycle time and fixed asymmetry factor $\eta$.  Figure~\ref{fig:workRIR}(b) then shows distributions for the work in a given cycle.  The average gives an estimate of the mean work at $\tau=1.25$.  Finally, fig.~\ref{fig:workRIR}(c) collects the mean-work estimates for different cycle times and $\eta~=$~4.  The solid lines are fits to the expected asymptotic form $\left <W\right >_{\tau} \sim \left<W\right>_{\infty} +a\tau^{-1}$ \cite{sekimoto97a}.  If work is measured in units of $kT$ and $\tau$ is dimensionless, then $a$ is a protocol-dependent constant of order unity.

\begin{figure}[h]
	 \begin{center}
	 \includegraphics[width=7cm]{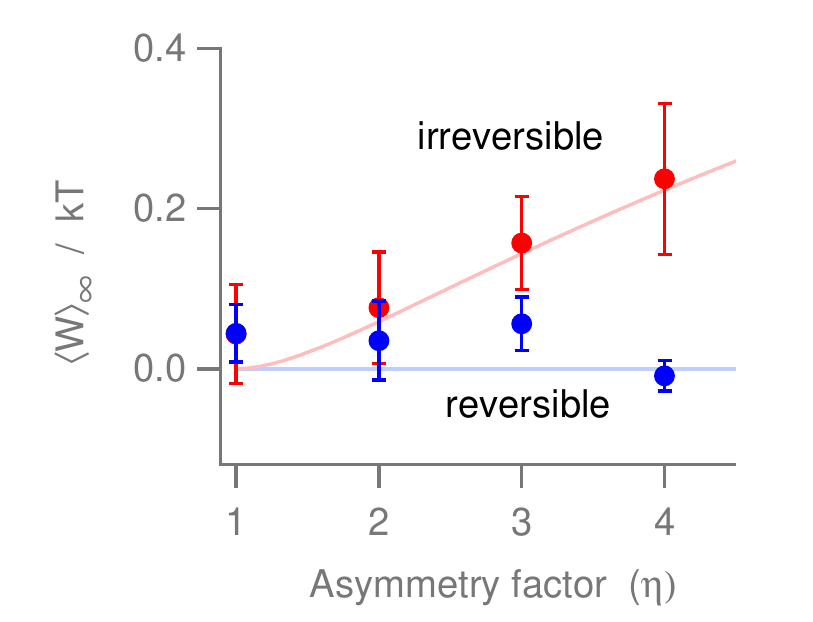}
	 \end{center}
	 \caption { \label{fig:AsymptoticWork} 
(Color online)  Asymptotic work as a function of asymmetry factor $\eta$.  The irreversible protocol leads to higher asymptotic work values than the reversible protocol for any $\eta\neq1$, in accordance with eq.~\ref{eq:WorkPrediction1}, plotted in light red.  The error bars are derived from least-squares fits to average work vs. inverse cycle time. The plot represents about eleven days of data. 
}
\end{figure}

By extrapolating the work to infinite cycle times ($\tau^{-1}~\to~0$), we find the asymptotic average work, $\left< W \right>_{\infty}$.  These values, as a function of $\eta$, are collected in fig.~\ref{fig:AsymptoticWork}.  The reversible protocol shows asymptotic mean work values consistent with 0, whereas the irreversible protocol shows that the work increases with $\eta$, even though the protocol is carried out arbitrarily slowly and returns the system to the equilibrium state.

\section{Discussion}
At first glance, one might think that both protocols should require no work when performed arbitrarily slowly and ostensibly quasistatically.  As we emphasized above, the protocols are nearly identical, differing only in the order of the first two operations (lowering the barrier and stretching the well, or vice versa).  Further, they both start from equilibrium and return to the same equilibrium state.  Yet the ``reversible" protocol requires no work in the $\tau^{-1}\to 0$ limit, whereas the ``irreversible" protocol does require work, even in the arbitrarily slow limit.  Why the difference?

By contrast, in the reversible protocol, the symmetry of the wells implies no net probability flux as the barrier is lowered.  In this case, the system is always very close to global thermodynamic equilibrium, and the transformation is reversible.
\begin{figure}[h]
	 \begin{center}
	 \includegraphics[width=7cm]{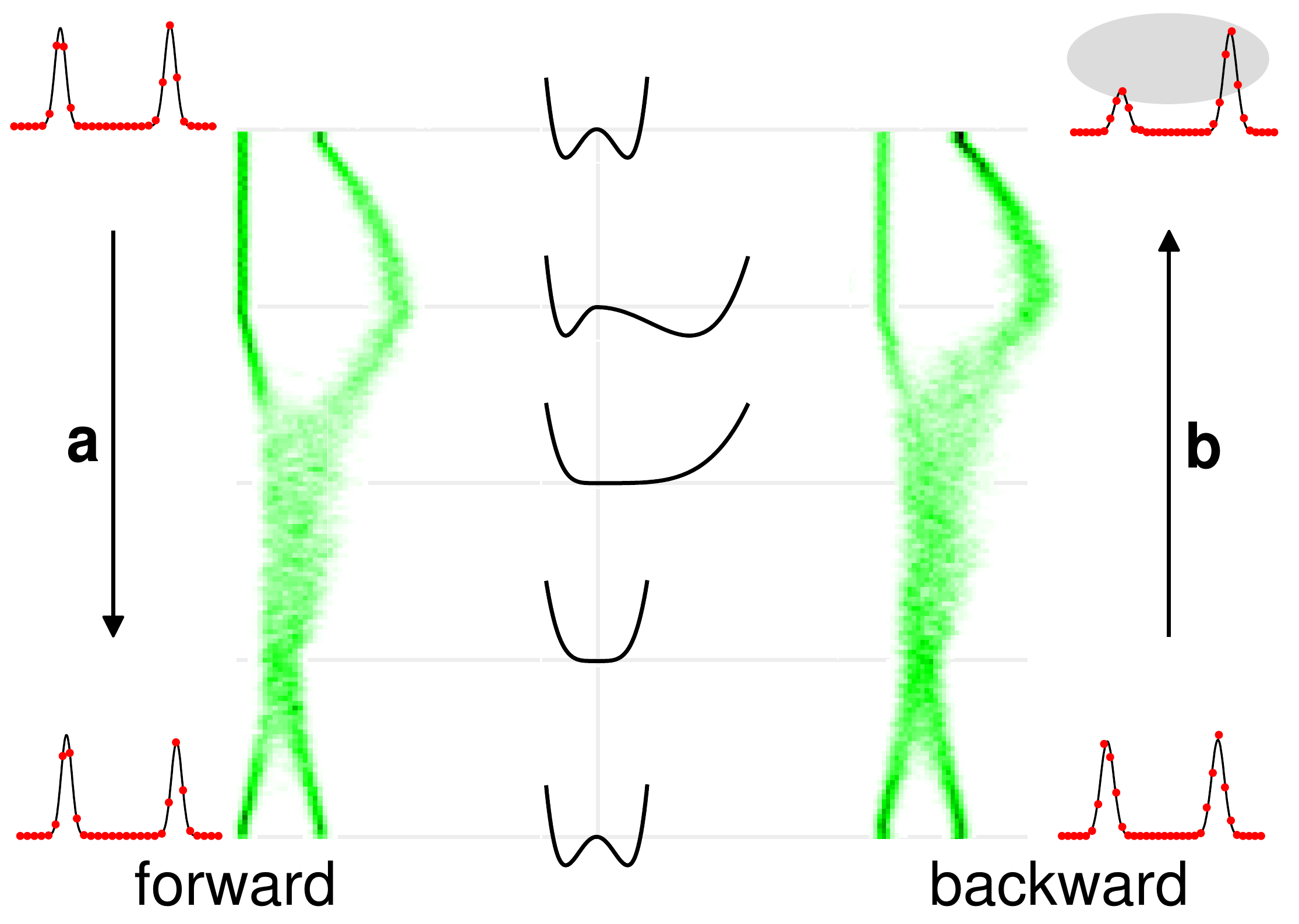}
	 \end{center}
	 \caption { \label{fig:ForBackProb} 
(Color online)  Transformation protocol and time evolution of probability in experiment recorded in forward and backward direction for fixed cycle time.  (a) The forward protocol has identical initial and final probabilities.  (b) The probability that a particle ends in the right well (upper right) is higher in the backward (time-reversed) protocol, as indicated in eq.~\ref{eq:WorkPrediction1}. The direction of time is indicated by the arrows at the edge.  Histograms show the spatial distribution of particles at the first and last time steps of the protocol.}
\end{figure} 

 To calculate the work required in the irreversible case, we can use recently developed tools of stochastic thermodyamics \cite{sekimoto10,seifert12}.  In particular, following the lead of Crooks \cite{crooks99}, we can compare the ``forward" experimental trajectories with hypothetical ``backwards" versions that result from reversing the protocol of potential manipulations.  When the reversible protocol, fig.~\ref{fig:workRIR}(b), is played backwards, a particle ends up in the left or right well with equal probability.  The final probabilities are then identical for both the forward and backward protocols.  But when the irreversible protocol, fig.~\ref{fig:workRIR}(a), is played backwards, the probabilities to end up in the left and right wells are not equal.  
 
We demonstrate the difference between forward and backward protocols experimentally by reversing the protocol for the irreversible transformation.  Figure~\ref{fig:ForBackProb} shows that the final probability of the reverse protocol \textit{differs} from the initial probabilities for the forward protocol; this also applies to the gas analogy in fig.~\ref{fig:gasSimple}(b).  

When the outcomes of forward and backward protocols differ, we can apply the relation \cite{kawai07,parrondo09,roldan14}
\begin{equation}
	\frac{\left< W \right>}{kT} \geq \frac{\Delta F}{kT} 
		+  D_{KL}(p_{i} || \bar{p}_{i}) \,,
\label{eq:MutualEntropy}
\end{equation}
where $\Delta F$ represents the equilibrium free energy difference between the initial and final states.  The term $p_{i}$ represents the probability that the system is in macrostate $i$ at time $t=0$ and $\bar{p}_{i}$ is probability that the system ends in macrostate $i$ at time $t=\tau$ under the reverse protocol.  The term $D_{KL}(p_{i} || \bar{p}_{i})$ is the Kullback-Leibler divergence between the probability distributions $p_{i}$ and $\bar{p}_{i}$, and $i$ corresponds to the left ($L$) and right ($R$) wells.  As usual, each macrostate corresponds to many microstates---here, trajectories that end up somewhere the left or right well.  We also note that the inequality in Eq.~\ref{eq:MutualEntropy} becomes an equality only in the arbitrarily slow limit.  At finite cycle times, more work, on average, will be required.

In our experiment, $\Delta F = 0$ for both protocols:  the potential is the same and the system is in thermodynamic equilibrium at the beginning and end of the protocols.  Further, since the arbitrarily slow limit gives a work approximately equal to that in the true quasistatic limit, the inequality becomes an equality.  In that limit, eq.~\ref{eq:MutualEntropy} simplifies to
$\left< W \right>/kT  
		= D_{KL}(p_i || \bar{p}_i) 
		= \sum p_i \, \ln \frac{p_i}{\bar{p}_i} $
, where $i=L$~or~$R$.
We set the initial probability for the particle to be in the left or right well to $p_{L} =p_{R} =1/2$ for both protocols.  The irreversible protocol is only in \textit{local} equilibrium when played forwards; however, it remains in global equilibrium when played backwards, except for the brief portion of the cycle where the energy barrier (while being lowered or raised) has a height such that the hopping time is comparable to the cycle time.  Since both protocols are in global equilibrium when played backwards and since both are performed quasistatically, we can use the Boltzmann distribution to estimate the probability that a particle ends up in the left or right well in the backwards protocols.  By integrating the position distribution for the potential in eq.~\ref{eq:DWpotential}, 
we find $\bar{p}_R = \eta \, \bar{p}_L$.  Normalizing $\bar{p}_R  + \bar{p}_L=1$ implies that $\bar{p}_L = 1/((1+\eta)$ and $\bar{p}_R = \eta/(1+\eta)$.
Note that we have first integrated out the $y$ coordinate in eq.~\ref{eq:DWpotential}, as that part of the potential is static and hence plays no role in calculating the work.  

For the reversible protocol, $\bar{p}_{L}=\bar{p}_{R}=1/2$, and the average work is $\left< W \right>~=~0$, as shown by the solid blue line in fig.~\ref{fig:workRIR}(c).  

But for the irreversible case, the work 
is
\begin{align}
	\frac{\left< W \right>}{kT} 
		&= \frac{1}{2} \left[ \ln \left( \frac{1/2}{1/(\eta+1)} \right) 
			+ \ln \left( \frac{1/2}{\eta / (\eta+1)} \right) \right] \nonumber \\
		&= \ln \left ( \frac{\eta+1}{2\sqrt{\eta}} \right) \,,
\label{eq:WorkPrediction1}
\end{align}
which gives the solid red line traced in fig.~\ref{fig:workRIR}(c).

\section{Conclusions}
Arbitrarily slow processes can be irreversible, when they involve mixing of two states that are not in equilibrium.  Here, we have shown that even when a cyclic transformation starts and ends in the same equilibrium state, the same may not be true of its time-reversed process.  In such cases, the transformation will require work to perform, on average, no matter how slowly it is carried out in the laboratory.  Although asymmetries between forward and backward processes capture the fundamental concept of irreversibility, the gas-piston analogy adds intuitive understanding:  a pressure difference is equalized by letting gas from the high-pressure compartment slowly bleed into the low-pressure compartment.

The analogy with an ideal gas in an ideal piston clarifies the conceptual origin of the irreversibility.
In both cases, we impose an internal constraint (the dividing wall, the high potential barrier) and then manipulate the two subsystems independently by breaking the left-right symmetry of the potential \cite{roldan14}.  When the internal constraint is removed, the transformation is irreversible if the two subsystems are not in global equilibrium (same pressure for a gas, same occupation probability for the mesoscopic system).
Remarkably, in a mesoscopic system, we can measure and demonstrate such effects, which otherwise have only been thought experiments for idealized macroscopic systems.  
Moreover, the mesoscopic version is subtler:  unlike the gas piston, whose pressures equalize with a ``whoosh," its nonequilibrium nature is not apparent in any one cycle but emerges only after the statistical analysis of many cycles.  And, because the hop time grows so rapidly with barrier height, smooth, ``innocent" parameter changes correspond to the more obvious opening and closing of classical valves.  Despite such subtleties, the experimental study of thermodynamics of small systems is not only useful for clarifying the role of fluctuations relative to the mean behavior, it is also, perhaps, the best way to approach ultimate thermodynamic limits.
 
\acknowledgments
We thank David Sivak, Karel Proesmans, Chris Van den Broeck, and Massimiliano Esposito for helpful discussions.  This work was supported by NSERC (Canada) and by a Billy Jones Graduate Award to MG.

\end{document}